\documentclass[aps,pra,twocolumn,showpacs,superscriptaddress]{revtex4}

\usepackage{amsfonts,mathrsfs,amsmath,amsthm,amssymb}
\usepackage{graphicx}
\usepackage{dcolumn}
\usepackage{bm,times}
\usepackage[colorlinks=true,breaklinks=true,linkcolor=blue,citecolor=blue,urlcolor=blue]{hyperref}

\def \tr{ {\rm{Tr}}}

\def \ket {\rangle}

\begin{document}

\title{Quantum discord for the general two-qubit case}

\author{Xiaohua Wu}
\email{wxhscu@scu.edu.cn}
\affiliation{College of Physical Science and Technology, Sichuan University, Chengdu 610064, China}
\author{Tao Zhou}
\email{taozhou@swjtu.edu.cn}
\affiliation{Quantum Optoelectronics Laboratory, School of Physics and Technology, Southwest Jiaotong University, Chengdu 610031, China}

\date{\today}

\begin{abstract}
Recently,   Girolami and Adesso have demonstrated that the calculation of quantum discord for two-qubit case can be viewed as to solve a pair of transcendental equation (Phys. Rev. A, {\bf 83}, 052108(2011)). In present work, we introduce  the generalized Choi-Jamiolkowski isomorphism and apply it as a convenient tool for constructing  transcendental equations. For the general two-qubit case, we show that the transcendental equations always have a finite set of universal solutions, this result can be viewed as a generalization of the one get by Ali, Rau, and Alber (Phys. Rev. A, {\bf 81}, 042105 (2010)).   For a subclass of $X$ state, we find the analytical solutions by solving   the transcendental equations.

\end{abstract}
 \maketitle

\section{introduction}
\label{intro}
How to quantify and characterize the nature of correlations in a quantum state, besides the fundamental scientific interest, has a crucial applicative importance in the field of quantum information processing~\cite{Nielsenbook}. For a bipartite quantum state, it is known that both the classical and quantum correlations are contained in it. Beyond the entanglement, quantum discord was introduced as a more general measure of quantum correlation~\cite{Ollivier,Henderson}, and was regarded as a resource for quantum computation~\cite{Datta}, quantum state merging~\cite{Madhok,Cavalcanti}. Quantum discord has attracted much attention recently~\cite{Datta,Madhok,Cavalcanti,Dakic,Luo,Luo2,Ali,Lu,Cen,Modi,Bellomo,discord}, and has also been generalized to continuous-variable systems for Gaussian states~\cite{Adesso,Giorda} and non-Gaussian states~\cite{Tatham}.

Quantum discord is very hard to calculate even for two-qubit states because of the minimization over all possible measurements. For an important class of two-qubit states, the so-called $X$ states, Ali, Rau, and Alber (ARA) proposed an algorithm to calculate the quantum discord with minimization taken over only a few simple cases ~\cite {Ali}. However, a counterexample  for the ARA algorithm was given by Lu \emph{et. al.}~\cite{Lu}, where the authors proved that, for the entire class os $X$ states, the optimization procedure involved in the classical correlation should be state dependent. For the real  $X$ states, Chen \emph{et. al.} have identified a class of states, where quantum discord can be evaluated analytically without any minimization, and hence the ARA algorithm is valid. Meanwhile, they also identified a family of states for which the ARA algorithm fails ~\cite{Chen}.

The ARA algorithm involved  a minimization procedure with four constrained parameters. However, Girolami and Adesso have shown that two free parameters, the polar and azimuthal angles usually used to describe an arbitrary unit Bloch vector, are already sufficient. With the two angles, one may obtain two partial derivatives for the conditional entropy, and by setting the two partial derivatives to be zero, the minimization procedure can be simplified as to find the solutions of a pair of transcendental equations~\cite{Girolami}. Usually, one should firstly give all the possible solutions, which are series of values of the two angles, and then select out the optimal setting where the conditional entropy takes the minimal value.

  Although the transcendental equations is direct and reliable, it has been argued that, for general case, one cannot solve the problem analytically since these equations involves logarithms of nonlinear quantities~\cite{Girolami}. In preset work, we shall give some further discussion about this problem. First, we introduce the generalized Choi-Jamiolkowski isomorphism as a convenient tool to construct the transcendental equations. Then, for the general two-qubit case, we demonstrate that the transcendental equations have a set of universal solutions which have been discovered by the ARA algorithm.   Finally, for a subclass of the $X$ state, we give the analytical solution by solving the transcendental equations.

The  content of present work is organized as follows. In Sec.~\ref{sec2}., we give a brief review of the quantum discord. In Sec.~\ref{sec3}., we introduce the general  Choi-Jamiolkowski isomorphism. In Sec.~\ref{Bloch}, a detail introduction of the Bloch vector transformation is discussed. In Sec.~\ref{setting}, we give a classification of the  solutions for the partial equation of the classical mutual information.  In Sec.~\ref{examples}, several examples are given there. Finally, we end our work with a short conclusion.

\section{The quantum discord}
\label{sec2}

The correlations for a bipartite state can be quantified by the quantum mutual information. For a given density matrix $\rho^{ab}$ of a bipartite system $H^a\otimes H^b$, the quantum mutual information is defined as
\begin{equation}
\mathcal{I}=S(\rho^a)+S(\rho^{b})-S(\rho^{ab}),
\end{equation}
where $S(\rho)=-\tr(\rho\log_2\rho)$ is the von Neumann entropy, and $\rho^a$ ($\rho^b$) denotes the reduced density matrix of subsystem $H^a$ ($H^b$). The quantum mutual information can be expressed as the sum of two part,
\begin{equation}
\mathcal{I}(\rho^{ab})=\mathcal{C}(\rho^{ab})+\mathcal {Q}(\rho^{ab}),
\end{equation}
with $\mathcal{C}(\rho^{ab})$ the classic correlation and $\mathcal {Q}(\rho^{ab})$ the quantum discord~\cite{Ollivier,Henderson}. To quantify the quantum discord, Olliver
and Zurek~\cite{Ollivier} has suggested the use of von Neumann type measurements: $\{\Pi_i\}_{i=1}^{D}$, with $\Pi_i$ the one-dimensional projective operators. After the measurement on subsystem $H^b$, a density operator $\rho_j$ associated with the outcome $j$ is
\begin{equation}
\rho_j=\frac{1}{p_j}(\mathrm{I}_D\otimes\Pi_j)(\rho^{ab})(\mathrm{I}_D\otimes\Pi_j),
\end{equation}
with $p_j$ the probability for the $j$-th outcome. Use $ S(\rho\vert{\Pi_j})=\sum p_j S(\rho_j)$ to denote the quantum conditional  entropy, and the corresponding quantum mutual information reads
\begin{equation}
\mathcal{I'}(\rho^{ab}\vert \Pi_j)=S(\rho^{a})-S(\rho\vert{\Pi_j}).
\end{equation}
Then, the classical correlation is
\begin{equation}
\label{equ6}
\mathcal{C}(\rho^{ab}):= \mathrm{sup}_{\{\Pi_j\}} \mathcal{I'}(\rho^{ab}\vert \Pi_j),
\end{equation}
and the quantum discord is defined as
\begin{equation}
\mathcal{Q}(\rho^{ab}):= \mathcal{I}(\rho^{ab})-\mathcal{C}(\rho^{ab}).
\end{equation}

\section{ The system-ancilla-environment picture}
\label{sec3}
The Choi-Jamiolkowski isomorphism is a useful connection between quantum channel and a bipartite state~\cite{Choi}, say, $\rho^{ab}=\varepsilon\otimes \mathrm{I}_D(\vert S_{+}\rangle \langle S_+\vert)$, with $\vert S_+\rangle$ the maximally entangled state of the bipartite system. Our work is motivated by such a simple idea: We first express a density operator $\rho^{ab}$ with a quantum channel $\varepsilon$, and then the analytic expression of the quantum discord for the $D\otimes D$ system can be simplified since only a $D$-dimensional quantum process $\varepsilon$ is involved. It should be noticed that the  isomorphism above is only available for the cases when the reduced density matrix $\rho^{b}$ is a completely mixture, $\rho^{b}=\mathrm{I}_D/D$. With careful analysis, we find that the isomorphism above can take a general form as the maximally entangled state is substituted by a general entangled state, and then, the density matrix with a full-rank reduced matrix $\rho^b$ can always be expressed with a quantum channel and an entangled state. Meanwhile, the Bloch vector transformation can be applied to describe the quantum operation for the qubit case. Therefore, the derivation of the quantum discord is closely related to the property of the quantum channel.

To study a quantum channel $\varepsilon$ of a $D$-dimensional system $H^a$, it is convenient to introduce an ancilla system $H^b$ with an equal dimension. Prepare a pure entangled state $|\Phi\ket$ as the initial state of the bipartite system $H^a\otimes H^b$, and since the system $H^a$ is subjected to a interaction described by the trace-preserving quantum operation $\varepsilon$ with the environment, the final state is
\begin{equation}
\label{isom}
\rho^{ab}=\varepsilon\otimes \mathrm{I}_D(\vert\Phi\rangle\langle\Phi\vert).
\end{equation}
From it, we can obtain a lot of information about the quantum channel. For example, the Schmacher's channel fidelity is defined as $F=\langle\Phi\vert\rho^{ab}\vert\Phi\rangle,$ which provides a measure of how well the entanglement between the two system is preserved by the quantum process $\varepsilon$~\cite{Schumacher}. In the following, we shall show that this process is reversible: If the reduced matrix of $H^b$ is full-rank, it can always be described by the corresponding $\varepsilon$ and $\vert\Phi\rangle$.
To prove this, we should first notice that a bounded operator in $\mathrm{H}_{D}$ is always related to a  vector in a extended Hilbert space $\mathrm{H}_{D}^{\otimes 2}$. Denote $A$ to be a bounded operator on the $D$-dimensional Hilbert space $\mathrm{H}_D$, with $A_{ij}=\langle i\vert  A\vert j\rangle$
the matrix elements, and an isomorphism between  $A$ and a $D^2$-dimensional vector $\vert A\rangle\rangle$ can be
 \begin{equation}
 \vert A\rangle\rangle =\sqrt{D} A\otimes \mathrm{I}_{D}\vert S_+\rangle=\sum_{i,j=1}^D A_{ij}\vert ij\rangle,
  \end{equation}
where $\vert S_+\rangle$ is the maximally entangled state in $\mathrm{H}_{D}^{\otimes 2}$, $\vert S_+ \rangle =\sum_{k=1}^{D}\vert kk\rangle/\sqrt{D}$ with $\vert ij\rangle=\vert i\rangle\otimes \vert j\rangle$.
This isomorphism offers a one-to-one map between an operator and its vector form. Suppose that  $A$ , $B$, and $\rho$  are three arbitrary bounded operators on $\mathrm{H}_{D}$, and then
\begin{equation}
\label{trace}
\tr(A^{\dagger}B)=\langle\langle A\vert B\rangle\rangle,
\vert A\rho B\rangle\rangle =A\otimes B^{\mathrm{T}}\vert \rho \rangle\rangle,
\end{equation}
with $B^{\mathrm{T}}$ the transpose of $B$.

Now, consider a $D^2\times D^2$ density matrix,
\begin{equation}
\label{density}
\rho^{ab}=\sum_{m=1}^{D^2}\lambda_{m}\vert\Psi_m\rangle\langle\Psi_m\vert
\end{equation}
where $\vert\Psi_m\rangle$ are the normalized eigenvectors of the bipartite density operator $\rho^{ab}$, $\langle\Psi_m\vert\Psi_{n}\rangle=\delta_{mn}$, and $\lambda_m$ are the corresponding eigenvalues with $\sum_{m}^{D^2}\lambda_m=1$. Due to the isomorphism $\vert \Psi_m\rangle =\vert \Gamma_m\rangle\rangle$, the density matrix $\rho^{ab}$ can be also expressed as
\begin{equation}
\rho^{ab}=\sum_{m}\lambda_m\vert \Gamma_m\rangle\rangle\langle\langle \Gamma_m\vert,\nonumber
\end{equation}
and the transpose of the reduced density matrix $\rho^{b}$ can be derived as
\begin{equation}
\label{tran}
(\rho^{b})^{\mathrm{T}}=\sum_m\lambda_m\Gamma_m^{\dagger}\Gamma_m.
\end{equation}
A simple proof of Eq.~(\ref{tran}) is as following: With the equations in Eq.~(\ref{trace}) and Eq.~(\ref{density}), one can obtain $\langle\langle \Gamma_m\vert ki\rangle=\langle i\vert \Gamma_m^{\dagger}\vert k\rangle$, and $\langle kj\vert \Gamma_m\rangle\rangle =\langle k\vert\Gamma_m\vert j\rangle$. Plugging these results into the definition of the partial trace operation, we have
 \begin{eqnarray}
 \langle i\vert (\rho^{b})^{\mathrm{T}}\vert j\rangle&=& \langle j\vert (\rho^{b})\vert i\rangle \nonumber\\
&=&\sum_{k=1}^{D}\langle kj\vert \sum_m \lambda_m \vert \Gamma_m\rangle\rangle\langle\langle \Gamma_m\vert ki\rangle\nonumber\\
&=&\sum_m\sum_k\lambda_m \langle i\vert\Gamma_m^{\dagger}\vert k\rangle\langle k\vert \Gamma_m\vert j\rangle\nonumber\\
&=&\langle i\vert \sum_m\lambda_m\Gamma_m^{\dagger}\Gamma_m\vert j\rangle. \nonumber
\end{eqnarray}
\qed

For the two-qubit case, $D=2$, and one can further assume $\mathrm{det} (\rho^{b})^{\mathrm{T}})\neq 0$, and define
\begin{equation}
\sqrt{(\rho^{b})^{\mathrm{T}}}=U\Omega U^{\dagger},
\end{equation}
where
\begin{equation}
\Omega=\left(
\begin{array}{cc}
\cos \frac{\gamma}{2} & 0 \\
0 & \sin\frac{\gamma}{2} \\
\end{array}
\right),\nonumber
\end{equation}
with $U$ a $2\times 2$ unitary transformation. Furthermore, one can introduce a set of Kraus operators $\{E_m\}_{m=1}^{4}$ for the quantum process $\varepsilon$ as follows:
\begin{equation}
E_m=\sqrt{\lambda_m}\Gamma_m U\Omega^{-1}, \sum_m E_m^{\dagger}E_m=\mathrm{I}_2,
\end{equation}
and now, the density matrix $\rho^{ab}$ can be rewritten as
 \begin{eqnarray}
 \rho^{ab}=\sum_m E_{m}\otimes \mathrm{I}_2 \vert \Omega U^{\dagger}\rangle \rangle\langle \langle \Omega U^{\dagger}\vert E^{\dagger}_{m}\otimes \mathrm{I}_2 \nonumber\\
 =\sum_m E_{m}\otimes U^* \vert \Omega \rangle\rangle \langle \langle \Omega \vert E^{\dagger}_{m}\otimes (U^*)^{\dagger}.\nonumber
 \end{eqnarray}
It is obvious that a new basis for $H^b$ can be defined as
\begin{equation}
\vert 0\rangle=(U^*)^{\dagger}\vert 0\rangle, \vert 1\rangle=(U^*)^{\dagger}\vert 1\rangle,\nonumber
\end{equation}
and the relation in Eq.~(\ref{isom}) with the entangled state
$$\vert\Phi\rangle=\vert\Omega\rangle\rangle = \cos\frac{\gamma}{2}\vert 00\rangle+\sin\frac{\gamma}{2}\vert 11\rangle.$$
can be obtained.
In this picture, the two reduced density matrices are
\begin{equation}
\rho^b=\left(
         \begin{array}{cc}
           \cos^2\frac{\gamma}{2} & 0 \\
           0 & \sin^2\frac{\gamma}{2}  \\
         \end{array}
       \right),\rho^a=\varepsilon(\rho^b).
    \end{equation}
It should be mentioned that $(\rho^{b})^{T}$ has the same determinant as $\rho^b$, and furthermore, our method above can be easily generalized for the cases with the arbitrary dimension $D$.

In the following, we shall focus on the situation where a von Neumann  measurement is performed on subsystem $H^b$. Two free parameters, $\theta$ and $\phi$, can be used for the measurement operators $\Pi_i=\vert\psi_i\rangle\langle\psi_i\vert(i=1,2)$, where
\begin{eqnarray}
\vert\psi_1\rangle&=&\cos\frac{\theta}{2}\vert 0\rangle+\sin\frac{\theta}{2}\exp(i\phi)|1\ket,\\
\vert \psi_2\rangle&=&-\sin\frac{\theta}{2}\vert 0\rangle+\cos\frac{\theta}{2}\exp(i\phi)|1\ket.
\end{eqnarray}
After the measurement, the final state $\rho^{a'b'}$ can be formally expressed as $\rho^{a'b'}=\varepsilon\otimes \mathrm{I }_2(\bar{\rho})$, where
\begin{equation}
\bar{\rho}=\sum_{j=1}^2\mathrm{I}_2\otimes \Pi_j\vert\Omega\rangle\rangle\langle\langle \Omega\vert \mathrm{I}_2\otimes \Pi_j.
\end{equation}
By some algebra, we find that $\bar{\rho}$ is a mixture of product state
\begin{equation}
\bar{\rho}=\sum_{j=1}^2 p_j\vert\phi_j\rangle \langle \phi_j\vert \otimes \vert\psi_j\rangle\langle \psi_j\vert,
\end{equation}
with $p_j$ the probabilities
\begin{equation}
\label{equ2}
p_1=\frac{1}{2}(1+\cos\theta\cos\gamma), p_2=\frac{1}{2}(1-\cos\theta\cos\gamma),
\end{equation}
and $\vert\phi_j\rangle (j=1,2)$ are a pair of pure states defined as
\begin{eqnarray}
\vert \phi_1\rangle&=&\frac{1}{\sqrt{p_1}}(\cos\frac{\gamma}{2}\cos\frac{\theta}{2}\vert 0\rangle+\sin\frac{\gamma}{2}\sin\frac{\theta}{2}e^{-i\phi}\vert 1\rangle),\nonumber\\
\vert \phi_2\rangle&=&\frac{1}{\sqrt{p_2}}(-\cos\frac{\gamma}{2}\sin\frac{\theta}{2}\vert 0\rangle+\sin\frac{\gamma}{2}\cos\frac{\theta}{2}e^{-i\phi}\vert 1\rangle).\nonumber
\end{eqnarray}
Finally, one can obtain
\begin{equation}
\label{densitry2}
\rho^{a'b'}=\sum_{j=1}^{2}p_j\rho_j\otimes \vert\psi_j\rangle\langle\psi_j\vert, \rho_j=\varepsilon(\vert\phi_j\rangle\langle\phi_j\vert).
\end{equation}
Meanwhile, it is easy to check that $\sum_jp_j\vert\phi_j\rangle\langle\phi_j\vert=\rho^b$, and therefore
\begin{equation}
\sum_{j=1}^2p_j\rho_j=\varepsilon(\rho^b)=\rho^a.
\end{equation}

From Eq.~(\ref{densitry2}), we see that the classic information $\mathcal{I}'$ is a function of the free parameters $\theta$ and $\phi$,
\begin{equation}
\mathcal{I}'(\theta,\phi)=S(\rho^a)-\sum_{j=1}^2p_jS(\rho_j),
\end{equation}
and the quantum discord can be accessed if the maximum value of $\mathcal{I}'(\theta,\phi)$ has been decided
\begin{equation}
\mathcal{Q}=\mathcal{I}-\mathrm{Max}_{\theta,\phi}\mathcal{I}'(\theta,\phi).
\end{equation}

\section{the Bloch vector transformation}
\label{Bloch}
In order to obtain the analytic expression of the quantum discord for a two-qubit state, we shall at first give a general expression of the conditional entropy $\sum_{j=1}^2p_jS(\rho_j)$. The Bloch representation is very useful for the single-qubit state, and the state $\rho$ can be written as $\rho=\frac{1}{2}(\mathrm{I}_2+ \vec{r}\cdot\vec{\sigma})$ with $\vec{r}$ is a three component real vector and $\vec{\sigma}=(\sigma_x,\sigma_y,\sigma_z)$. Meanwhile, it turns out that an arbitrary trace-preserving quantum operation is equivalent to a map such that
\begin{equation}
\vec{r'}\rightarrow\vec{r}=\eta\vec{r}+\vec{c},
\end{equation}
with $\eta$ a $3\times 3$ real matrix, $\vec{c}$ a constant vector, and  $\varepsilon(\rho)=\frac{1}{2}(\mathrm{I}_2+ \vec{r'}\cdot\vec{\sigma})$. This is an affine map, mapping the Bloch sphere into itself~\cite{Nielsenbook}, and can be explicitly expressed as
\begin{equation}
\label{map}
\left(
  \begin{array}{c}
    r'_x \\
    r'_y \\
    r'_z \\
  \end{array}
\right)=\left(
          \begin{array}{ccc}
            \eta_{xx} & \eta_{xy}& \eta_{xz} \\
            \eta_{yx} & \eta_{yy} & \eta_{yz} \\
            \eta_{zx} & \eta_{zy} & \eta_{zz} \\
          \end{array} \right)
          \left(
                       \begin{array}{c}
                         r_x \\
                         r_y \\
                         r_z \\
                       \end{array}
                     \right)+\left(
                               \begin{array}{c}
                                 c_x \\
                                 c_y\\
                                 c_z\\
                               \end{array}
                             \right),
\end{equation}
with the coefficients defined as
\begin{equation}
\eta_{ij}=\frac{1}{2}\tr[\sigma_j\varepsilon(\sigma_i)],c_k=\frac{1}{2}\tr[\sigma_k\varepsilon(\mathrm{I}_2)].
\end{equation}
Here, we have used
\begin{equation}
\vert\phi_1\rangle\langle \phi_1\vert= \frac{1}{2}(\mathrm{I}_2+ \vec{s}\cdot\vec{\sigma}),\ \ \vert\phi_2\rangle\langle \phi_2\vert= \frac{1}{2}(\mathrm{I}_2+ \vec{t}\cdot\vec{\sigma}),
\end{equation}
and the two unit vectors $ \vec{s}$ and $\vec{t}$
\begin{eqnarray}
\label{equ1}
s_x&=&\frac{\sin\gamma\sin\theta\cos\phi}{1+\cos\gamma\cos\theta},
s_y=\frac{\sin\gamma\sin\theta\sin\phi}{1+\cos\gamma\cos\theta},\nonumber\\
s_z&=&\frac{\cos\gamma+\cos\theta}{1+\cos\gamma\cos\theta},
t_x=\frac{-\sin\gamma\sin\theta\cos\phi}{1-\cos\gamma\cos\theta},\\
t_y&=&\frac{-\sin\gamma\sin\theta\sin\phi}{1-\cos\gamma\cos\theta},
t_z=\frac{\cos\gamma-\cos\theta}{1-\cos\gamma\cos\theta}.\nonumber
\end{eqnarray}
With the following two vectors,
\begin{equation}
\vec{s}'=\eta\vec{s}+\vec{c},\vec{t}'=\eta\vec{t}+\vec{c}
\end{equation}
one may have
\begin{equation}
\rho_1=\frac{1}{2}(\mathrm{I}_2+ \vec{s}'\cdot\vec{\sigma}),\rho_2=\frac{1}{2}(\mathrm{I}_2+ \vec{t}'\cdot\vec{\sigma}),
\end{equation}
For simplicity, $s'(\theta,\phi)$ and $t'(\theta, \phi)$ are used to denote the purity of the density matrix $\rho_1$ and $\rho_2$ respectively, and $s'(\theta,\phi)=\vert \vec{s}'\vert=\sqrt{(s'_x)^2+(s'_y)^2+(s'_z)^2}$, $ t'(\theta,\phi)=\vert \vec{t'}\vert$. It is easy to note that there exists a symmetry between these two functions: Under the transformation
 \begin{equation}
 \theta\rightarrow \pi-\theta, \phi\rightarrow \phi+\pi
 \end{equation}
 these two functions are interchanged
 \begin{equation}
 s'(\theta,\phi)\Longleftrightarrow t'(\theta,\phi).
 \end{equation}
 This result comes from the fact that $\vec{r}(\pi-\theta, \phi+\pi)=\vec{s}(\theta, \phi)$, which can be seen from Eq.~(\ref{equ1}).

\section { classification of the solutions}
\label{setting}
 With the Bloch vector introduced in above section, one may get a general expression of the classic information,
 \begin{equation}
 \mathcal{I}'(\theta,\phi)=S(\rho^a)-p_1H_2(\frac{1+s'}{2})-p_2H_2(\frac{1+t'}{2}),
 \end{equation}
 with $H_2(p)$ the binary entropy defined as $H_2(p)=-p\log_2 p-(1-p)\log_2(1-p)$.
From Eq.~(\ref{equ2}), $\frac{\partial p_1}{\partial \theta}=-\frac{\partial p_2}{\partial\theta}$. Therefore, we can obtain
\begin{eqnarray}
\frac{\partial \mathcal{I}'}{\partial \phi}&=&\frac{\partial s'}{\partial\phi}(p_1\log_2\sqrt{\frac{1+s'}{1-s'}})+\frac{\partial t'}{\partial\phi}(p_2\log_2\sqrt{\frac{1+t'}{1-t'}}),\nonumber\\
\frac{\partial \mathcal{I}'}{\partial \theta}&=&-\frac{\partial p_1}{\partial \theta}(H_2(\frac{1+s'}{2})-H_2(\frac{1+t'}{2}))\\
&&+\frac{\partial s'}{\partial\theta}(p_1\log_2\sqrt{\frac{1+s'}{1-s'}})+\frac{\partial t'}{\partial\theta}(p_2\log_2\sqrt{\frac{1+t'}{1-t'}}).\nonumber
 \end{eqnarray}
As a necessary condition, the maximum value may happen with
\begin{equation}
\label{tranequ}
\partial \mathcal{I}'/\partial \phi=0, \partial \mathcal{I}'/\partial \theta=0.
\end{equation}
In the following, we shall show that the  following two types of  solutions are universal:

(A)\emph{The symmetric solution}: In this case, one of the solutions happens with the setting
\begin{equation}
\label{equ3}
\theta=\frac{\pi}{2},\phi=\bar{\phi },
\end{equation}
with $\bar{\phi }$ is constrained by
\begin{eqnarray}
s'(\theta,\bar{\phi})=s'(\theta, \pi+ \bar{\phi }),\\
\frac{\partial s'(\pi/2, \phi)}{\partial \phi}\vert_{\phi=\bar{\phi}}=-\frac{\partial t'(\pi/2, \phi)}{\partial \phi}\vert_{\phi=\bar{\phi}}.
\end{eqnarray}

Following the discussions about the symmetry between $s'(\theta, \bar{\phi})$ and $t'(\theta, \bar{\phi})$, there should be
\begin{eqnarray}
s'(\pi/2,\bar{\phi})=t'(\pi/2,\bar{\phi}),\\
\frac{\partial s'(\theta, \bar{\phi})}{\partial \theta}\vert_{\theta=\pi/2}=-\frac{\partial t'(\theta, \bar{\phi})}{\partial \theta}\vert_{\theta=\pi/2}.
\end{eqnarray}
Jointing the above results with
\begin{equation}
p_1(\theta=\pi/2)=\frac{1}{2}, p_2(\theta=\pi/2)=\frac{1}{2},
 \end{equation}
 one can conclude that the setting in Eq.~(\ref{equ3}) is one of the solutions.

(B)\emph{The asymmetric solution}: Another  solution of the partial equation exists with the setting
\begin{equation}
\label{equas}
\theta=\tilde{\theta}, \phi=\tilde{\phi},
\end{equation}
with $\tilde{\theta}$ and $\tilde{\phi}$ the solution of the equations below,
\begin{eqnarray}
\frac{\partial s'(\theta, {\tilde{\phi}})}{\partial \theta}\vert_{\theta=\tilde{\theta}}=\frac{\partial t'(\theta, {\tilde{\phi}})}{\partial \theta}\vert_{\theta=\tilde{\theta}}=0,\\
\frac{\partial s'(\tilde{\theta}, \phi)}{\partial \phi}\vert_{\phi=\tilde{\phi}}=\frac{\partial t'(\tilde{\theta}, \phi)}{\partial \phi}\vert_{\phi=\tilde{\phi}}=0,\\
\frac{\partial p_1}{\partial \theta}\vert_{\theta=\tilde{\theta}}=\frac{\partial p_2}{\partial \theta}\vert_{\theta=\tilde{\theta}}=0,
\end{eqnarray}
Except the special case where $p_1=p_2=1/2$, $\theta =0$ is the only possible solution for the equations above since $\partial p_i/\partial \theta\propto\sin\theta,(i=1,2)$. By jointing it with the symmetric solution,
$\theta=\pi/2$,  the main result in~\cite{Ali}, which states that the polar angle $\theta$ may take the value $0$ or $\pi/2$ for the $X$ states, is also suitable for the general two-qubit case.

(C)\emph{The state-dependent solution}: For some given states, the two transcendental equations in Eq.~(\ref{tranequ}) may have other type solutions beside the universal one given above. We shall give an example in the next section.

\section{examples}
\label{examples}
 (A) \emph{The $X$ state.} This type of density matrix has been widely discussed in previous works~\cite{Luo,Ali,Lu},
 \begin{equation}
\rho^{ab}=\left(
            \begin{array}{cccc}
              \rho_{11} & 0 & 0 & \rho_{14} \\
              0 & \rho_{22} & \rho_{23} & 0 \\
              0 & \rho_{32} & \rho_{33} & 0 \\
              \rho_{41} & 0 & 0 & \rho_{44} \\
            \end{array}
          \right).
          \end{equation}

By some simple algebra, we may see that the map in Eq.~(\ref{map}) now take the form
\begin{equation}
\left(
  \begin{array}{c}
    r'_x \\
    r'_y \\
    r'_z \\
  \end{array}
\right)=\left(
          \begin{array}{ccc}
            \eta_{xx} & \eta_{xy}& 0 \\
            \eta_{yx} & \eta_{yy} & 0 \\
            0 & 0 & \eta_{zz} \\
          \end{array} \right)
          \left(
                       \begin{array}{c}
                         r_x \\
                         r_y \\
                         r_z \\
                       \end{array}
                     \right)+\left(
                               \begin{array}{c}
                                0 \\
                                 0\\
                                 c_z\\
                               \end{array}
                             \right),
\end{equation}

 Here, we focus on the case when $\cos\gamma=0$, which means the pure state $\vert\Phi\rangle$ in Eq.~(\ref{isom}) is the maximally entangled state. Under this condition, the probability for each final state takes the same value, $p_1=p_2=1/2$. The transcendental equations are reduced as
 \begin{eqnarray}
0&=&\frac{\partial s'}{\partial\phi}(\log_2\sqrt{\frac{1+s'}{1-s'}})+\frac{\partial t'}{\partial\phi}(\log_2\sqrt{\frac{1+t'}{1-t'}}),\\
0&=&\frac{\partial s'}{\partial\theta}(\log_2\sqrt{\frac{1+s'}{1-s'}})+\frac{\partial t'}{\partial\theta}(\log_2\sqrt{\frac{1+t'}{1-t'}}).
 \end{eqnarray}

With the  vector transformation, we shall get
\begin{eqnarray}
s'(\theta,\phi)&=&\{\sin^2\theta f(\phi)
+[c_z+\eta_{zz}\cos\theta]^2\}^{\frac{1}{2}},\\
t'(\theta,\phi)&=&\{\sin^2\theta f(\phi)
+[c_z-\eta_{zz}\cos\theta]^2\}^{\frac{1}{2}},\\
f(\phi)&=&(\eta_{xx}\cos\phi+\eta_{xy}\sin\phi)^2+(\eta_{yx}\cos\phi+\eta_{yy}\sin\phi)^2\nonumber
\end{eqnarray}
Note that  $s'$ and $t'$ depend on $\phi$ in the same way. Therefore, the optimal setting for $\phi$ should be decided by equation, $\partial f(\phi)/\partial\phi=0$, can be easily solved. By introducing a set of parameters,
\begin{eqnarray}
\eta_{\bot}^2&=&\max_{\phi}f(\phi), a=\eta_{\bot}^2+c_z^2, b=\eta_{zz}c_z,\\
c&=&\eta_{zz}^2-\eta^2_{\bot}, k=\frac{c}{b^2-ca},
\end{eqnarray}
we can express $s'$ and $t'$ with a simple form,
\begin{equation}
s'=\{a+2b\cos\theta+c\cos^2\theta\}^\frac{1}{2},t'=\{a-2b\cos\theta+c\cos^2\theta\}^\frac{1}{2}.\nonumber
\end{equation}
From it, we get the derivatives,
\begin{eqnarray}
\vert \frac{\partial s'}{\partial \theta}\vert&=& \sin\theta \sqrt{b^2-ca}G(s'),\nonumber\\
\vert \frac{\partial t'}{\partial \theta}\vert&=& \sin\theta \sqrt{b^2-ca}G(t'),\nonumber\\
G(x)&=&\frac{\sqrt{1+k x^2}}{x}, 0<x<1.
\end{eqnarray}
Note that  two derivatives can not be positive at the same time, we can rewrite Eq. (49) as
\begin{eqnarray}
0&=&\frac{\sin\theta \sqrt{b^2-ca}}{2\ln 2}(H(s')-H(t')),\\
H(x)&=&\frac{\sqrt{1+k x^2}}{x}\ln \frac{1+x}{1-x}, 0<x<1
\end{eqnarray}
Here, we shall show that: In the parameter range
\begin{equation}
k\geq-\frac{2}{3} ~or~ k\leq -1,
\end{equation}
the equation in (55) has no other solutions beside $\theta=0$ and $s'=t'$.  From Eq. (56), there should be
\begin{equation}
\frac{\partial H(x)}{\partial x}=-\frac{1}{x^2\sqrt{1+kx^2}}[\ln\frac{1+x}{1-x}-2x\frac{(1+kx^2)}{1-x^2}]
\end{equation}
with the expanding formula
\begin{eqnarray}
\ln\frac{1+x}{1-x}=2x(1+\sum_{n=1}\frac{x^{2n}}{2n+1}),\nonumber\\
\frac{(1+kx^2)}{1-x^2}=1+(1+k)\sum_{n=1}x^{2n},\nonumber
\end{eqnarray}
and  the condition in Eq. (57), we see that  $\partial H(x)/\partial x$ is non-zero in the parameter range $0<x<1$. Therefore, $H(s')=H(t')$ can only happen with $s'=t'$. If $c_z\neq 0$, $s'=t'$ has the unique setting $\theta=\pi/2$. If $c_z=0$, $s'=t'$ can hold for an arbitrary $\theta$, while from Eqs. (50-51), we find the optimal setting  is either  $\theta=0$ or $\theta=\pi/2$. Based on these analysis above, we conclude  that the universal solutions are sufficient for the cases above.

Among all the $X$-type states, the Bell diagonal state is one of the most interesting cases, and in the parameterized state model here, it corresponds to the situation
\begin{equation}
\left(
  \begin{array}{c}
    r'_x \\
    r'_y \\
    r'_z \\
  \end{array}
\right)=\left(
          \begin{array}{ccc}
            \eta_{xx} &0& 0 \\
            0 & \eta_{yy} & 0 \\
            0 & 0 & \eta_{zz} \\
          \end{array} \right)
          \left(
                       \begin{array}{c}
                         r_x \\
                         r_y \\
                         r_z \\
                       \end{array}
                     \right)
                               \end{equation}
Now, the parameter $k$ takes the value $k=-{1}/{\eta^2_{\bot}}<-1$.  The symmetric solution should be
$s'(\pi/2, \bar{\phi})=t'(\pi/2, \bar{\phi})=\eta_{\bot}$
while the asymmetric solution
 has a compact form
$s'(\tilde{\theta},\tilde{\phi})=t'(\tilde{\theta},\tilde{\phi})=\vert \eta_{zz}\vert$.
Finally, which kind of solution, the symmetric one or the asymmetric one, should be viewed as the classic correlation $\mathcal {C}$ in Eq.~(\ref{equ6}), is  decided by the actual values of $\eta_{xx}$, $\eta_{yy}$,and $\eta_{zz}$. Formally, it an be expressed as
\begin{equation}
\mathcal{C}=1-H_2(\frac{1+\eta_{\mathrm{opt}}}{2}),
\end{equation}
with $\eta_{\mathrm{opt}}=\mathrm{Max}\{\vert \eta_{xx}\vert,\vert\eta_{yy}\vert,\vert\eta_{zz}\vert\}$.

(B) In Ref. ~\cite{Lu}, a simple density matrix is given as
 \begin{equation}
\rho^{ab}=\left(
            \begin{array}{cccc}
              0.0783 & 0 & 0 & 0 \\
              0 & 0.1250 & 0.1000 & 0 \\
              0 & 0.1000 & 0.1250 & 0 \\
              0 & 0 & 0 & 0.6170 \\
            \end{array}
          \right).
          \end{equation}
With numerical calculation, we find the transcendental equations in Eq.~(\ref{tranequ}) have three solutions, $\theta=0$, $\theta=\pi/2$ and $\theta\approx 0.155\pi$. Among all these possible settings, $\theta\approx 0.155\pi$ is the optimal one. With this  simple example, we show that the universal solutions are not always the optimal one.

\section{Conclusions}
\label{conclusion}
Our present work has followed the original definition of the quantum discord in Ref.~\cite{Ollivier}, where the von Neumann projective measurement is performed. This measurement can also be generalized to the more general positive operator-valued measurement (POVM)~\cite{Henderson}. Furthermore, the concept of the quantum discord itself has been developed in recent years. For examples, the relative entropy quantum discord~\cite{Modi}, the geometric  quantum discord~\cite{Bellomo,Dakic} and their relations to the original definition have been investigated. Although our derivation is the for the original quantum discord, the general Choi-Jamiolkowski isomorphism used here may also be applied for the discussion for other types of quantum discord.

In summary, we have applied the general Choi-Jamiolkowski isomorphism  as a convenient tool for constructing the transcendental equations. For the general two-qubit case, we have shown that the transcendental equations always have a finite set of universal solutions, this result can be viewed as a generalization of the one get with the ARA algorithm. However, for some cases, the transcendental equations can have solutions beside the universal ones.  We also consider a subclass of $X$ state, for which the transcendental equation may offer
analytical solutions.

\section*{Acknowledgements}
This work was partially supported by the National Natural Science Foundation of China under the Grant No.~11405136, and the Fundamental Research Funds for the Central Universities of China A0920502051411-56.

\end{document}